\documentclass{mem}
\usepackage{natbib}\usepackage{txfonts}\usepackage{balance}
\usepackage{graphicx}
\idline{00}{0}
\usepackage{txfonts} 

\begin{document}
\def\teff{$T\rm_{eff }$}
\def\kms{$\mathrm {km s}^{-1}$}

\title{Basic properties of Narrow-Line Seyfert 1 Galaxies with Relativistic Jets}

\author{L. Foschini\inst{1},
E. Angelakis\inst{2},
G. Bonnoli\inst{1}, 
V. Braito\inst{1},
A. Caccianiga\inst{1}
L. Fuhrmann\inst{2},
L. Gallo\inst{3},
G. Ghirlanda\inst{1},
G. Ghisellini\inst{1},
D. Grupe\inst{4}, 
T. Hamilton\inst{5},
S. Kaufmann\inst{6},
S. Komossa\inst{2},
Y.~Y. Kovalev\inst{7,2},
A. Lahteenmaki\inst{8}, 
M. L. Lister\inst{9}, 
K. Mannheim\inst{10},
L. Maraschi\inst{1}, 
S. Mathur\inst{11},
B.~M. Peterson\inst{11},
P. Romano\inst{12},
P. Severgnini\inst{1},
G. Tagliaferri\inst{1},
J. Tammi\inst{8},
F. Tavecchio\inst{1},
O. Tibolla\inst{10},
M. Tornikoski\inst{8},
S. Vercellone\inst{12}
}

\institute{
INAF -- Osservatorio Astronomico di Brera, Milano/Merate, Italy\\
\email{luigi.foschini@brera.inaf.it}
\and
Max-Planck-Institut f\"ur Radioastronomie, Bonn, Germany
\and
Department of Astronomy and Physics, Saint Mary's University, Halifax, Canada
\and
Department of Astronomy \& Astrophysics, Pennsylvania State University, USA
\and
Department of Natural Sciences, Shawnee State University, Portsmouth, USA
\and
Landessternwarte University of Heidelberg, Heidelberg, Germany
\and
Astro Space Center of the Lebedev Physical Institute, Moscow, Russia
\and
Aalto University Mets\"ahovi Radio Observatory, Kylm\"al\"a, Finland 
\and
Department of Physics, Purdue University, West Lafayette, USA
\and
ITPA, Universit\"at W\"urzburg, W\"urzburg, Germany
\and
Astronomy Department, Ohio State University, Columbus, USA
\and
INAF -- IASF, Palermo, Italy
}

\authorrunning{Foschini et al.}

\titlerunning{NLS1s with relativistic jets}

\abstract{We present the preliminary results of a survey performed with {\it Swift} to observe a sample of radio-loud Narrow-Line Seyfert 1 Galaxies (RLNLS1s). Optical-to-X-ray data from {\it Swift} are complemented with $\gamma$-ray observations from {\it Fermi}/LAT and radio measurements available in the literature. The comparison with a sample of bright {\it Fermi} blazars indicates that RLNLS1s seem to be the low-power tail of the distribution.
}
\maketitle{}

The recent discovery by {\it Fermi}/LAT of high-energy $\gamma$-ray emission from some radio-loud narrow-line Seyfert 1 Galaxies (RLNLS1s, Abdo et al. 2009) has drawn significant attention to this rather poorly known class of active galactic nuclei (AGN). Our understanding of these sources is hampered by the fact that multi-wavelength archives contain scarce, sparse, and not simultaneous information about these sources. Early attempts to characterize the main properties of RLNLS1s were performed on one-two dozens of sources on the basis of radio-to-X-ray data (Whalen et al. 2006, Komossa et al. 2006, Yuan et al. 2008). The first radio-to-$\gamma-$ray study of RLNLS1s has been presented by Foschini (2011): it included 76 NLS1s (46 radio-loud, 30 radio-quiet as a control sample) and 34 quasars. Radio data were extracted from VLA FIRST survey at 1.4 GHz, while optical B filter magnitudes were from the {\it SDSS} or {\it USNO B1}. {\it ROSAT} provided X-ray information at 1 keV (60\% detections), while $\gamma$ rays were from {\it Fermi}/LAT (30 months of data, IRF P6). 

\begin{figure}[t!]
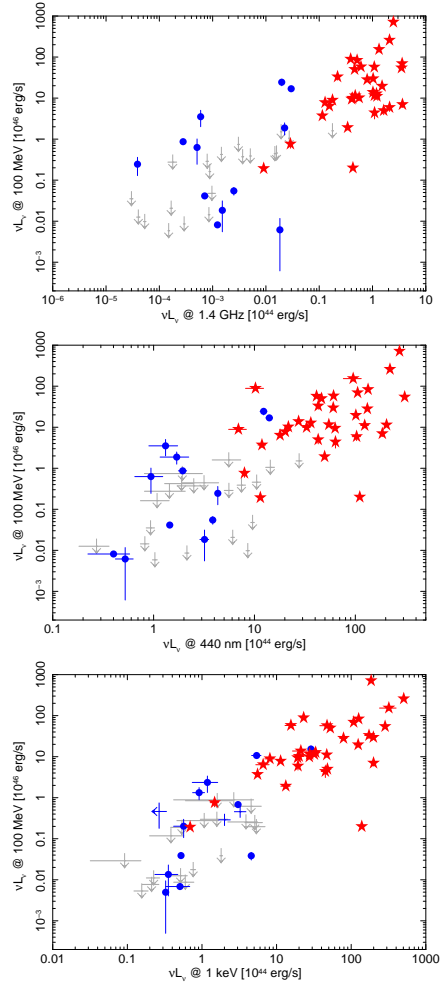

\includegraphics[angle=270,scale=0.23]{foschinil_f1.ps}\\
\includegraphics[angle=270,scale=0.23]{foschinil_f2.ps}\\
\includegraphics[angle=270,scale=0.23]{foschinil_f3.ps}\\
\vskip 6pt
\caption{\footnotesize $\gamma$-ray emission (Y axes) vs radio (1.4 GHz, top panel), optical (440 nm, middle panel), and X-ray (1 keV, bottom panel) emission for the sample of quasars (red stars) and RLNLS1s (blue circles or upper limits). The values have been $K$-corrected by using typical spectral indexes.}
\label{graphs}
\end{figure}

To improve the X-ray detection rate, we performed a series of snapshots with {\it Swift}. The number of known RLNLS1s is today relatively small (49 sources\footnote{See: {\tt http://tinyurl.com/gnls1s}}) and therefore it was possible to obtain a complete coverage with a reasonable amount of observing time. Fig.~\ref{graphs} shows the preliminary results of the observations performed to date. About 65\% of the sources have been observed, with a X-ray detection rate with {\it Swift}/XRT of 91\%. The optical B filter fluxes have been extracted from {\it Swift}/UVOT, while {\it Fermi}/LAT data are now derived from the analysis with IRF P7 of 46 months of observations. 

RLNLS1s smoothly overlap the quasar region, extending toward the low-power fluxes. The basic immediate information coming from these graphs is that the RLNLS1s seem to be the low-power part of the distribution. This is somehow expected, since the mass of the central black hole of NLS1s is lower than quasars: indeed, once renormalized for the mass, the jet power of RLNLS1s is comparable with that of quasars (see Foschini 2012).

The work is in progress to include other data (high-frequency radio core fluxes, ultraviolet magnitudes) and further details will be published in a comprehensive paper in preparation.

\bibliographystyle{aa}

\end{document}